\documentclass[copyright,creativecommons]{eptcs}
\providecommand{\event}{ICE 2012} 

\usepackage{amsmath}
\usepackage{amssymb}
\usepackage{amsfonts}
\usepackage{enumerate}
\usepackage{times}
\usepackage{bm}
\usepackage{tikz}
\usepackage[all]{xy}
\usepackage{multirow}
\usepackage{algorithm}
\usepackage{algorithmic}
\usepackage{hyperref}
\usepackage{amsthm}
\usepackage{stmaryrd}
\usepackage{enumerate}
\usepackage{bm}

\newcommand{\tellp}[1]{\tell(#1)}
\newcommand{\askp}[2]{\ask \  #1 \  \rightarrow \ #2}

\newcommand{\true}{{\it true}}
\newcommand{\false}{{\it false}}

\newcommand{\ask}{{\bf ask}}

\newcommand{\tell}{{\bf tell}}
\newcommand{\Stop}{{\bf stop}}

\newcommand{\rrarrow}{\longrightarrow}

\newcommand{\Con}{\mathit{Con}}

\newcommand{\pairccp}[2]{\langle #1,#2 \rangle}
\newcommand{\trans}[1]{\stackrel{#1}{\rrarrow}}
\newcommand{\tr}[1]{\trans{#1}}

\newcommand{\truep}{\mbox{\true}}
\newcommand{\stopp}{\mbox{\Stop}}
\newcommand{\barb}[1]{\downarrow_{#1}}
\newcommand{\wbarb}[1]{\Downarrow_{#1}}

\newcommand{\satbis}{\dot{\sim}_{sb}}
\newcommand{\wsatbis}{\dot{\approx}_{sb}}

\newcommand{\satstbisim}{\dot{\sim}_{sb}}
\newcommand{\bigfrac}[2]{
\begin{array}{c}#1\\
\hline #2
\end{array}}

\newtheorem{theorem}{Theorem}
\newtheorem{proposition}{Proposition}
\newtheorem{definition}{Definition}
\newtheorem{remark}{Remark}
\newtheorem{corollary}{Corollary}
\newtheorem{lemma}{Lemma}
\newtheorem{example}{Example}
\newtheorem{notation}{Notation}

\newcommand{\newtrans}[1]{\stackrel{#1}{\Longrightarrow}}
\newcommand{\newreds}{\Longrightarrow^*}
\newcommand{\reds}{\rrarrow^*}
\newcommand{\gentrans}[1]{\stackrel{#1}{\rightsquigarrow}}

\newcommand{\genbarb}[1]{\lightning_{#1}}

\newcommand{\R}{\mathcal{R}}
\newcommand{\A}{\alpha}
\newcommand{\B}{\beta}
\newcommand{\C}{\lambda}
\newcommand{\G}{\gamma}
\newcommand{\conf}[2]{\pairccp{#1}{#2}}
\newcommand{\deriv}{\vdash_D}
\newcommand{\derivR}{\vdash_{\R}}

\newcommand{\dom}{\succ_D}
\newcommand{\domR}{\succ_{\R}}

\newcommand{\transition}[5]{\conf{#1}{#2} \trans{#3} \conf{#4}{#5}}
\newcommand{\redstransition}[4]{\conf{#1}{#2} \reds \conf{#3}{#4}}
\newcommand{\newtransition}[5]{\conf{#1}{#2} \newtrans{#3} \conf{#4}{#5}}

\newcommand{\lub}{\sqcup}
\newcommand{\entailed}{\sqsubseteq}
\newcommand{\Conf}{\mathit{Conf}}
\newcommand{\idrel}{\mathit{id}}
\newcommand{\irrbis}{\dot{\sim}_{I}}
\newcommand{\symbis}{\dot{\sim}_{sym}}
\newcommand{\rel}[4]{\conf{#1}{#2} \R \conf{#3}{#4}}

\newcommand{\rTau}{{\sf R-Tau} }
\newcommand{\rLabel}{{\sf R-Label} }
\newcommand{\rAdd}{{\sf R-Add} }
\newcommand{\newirrbis}{\dot{\sim}^{\newtrans{}}_{I}}
\newcommand{\newsymbis}{\dot{\sim}^{\newtrans{}}_{sym}}

\newcommand{\nonewirrbis}{\not \!\!\! \newirrbis}

\title{Reducing Weak to Strong Bisimilarity in CCP
\footnote{This work has been partially supported by the project ANR-09-BLAN-0169-01 PANDA,
and by the French Defence procurement agency (DGA) with two PhD grants.}}
\author{
Andr\'es Aristiz\'abal \institute{CNRS/DGA and LIX \'Ecole Polytechnique de Paris}
\and
Filippo Bonchi 
\institute{ENS Lyon, Universit\'e
de Lyon, LIP (UMR 5668 CNRS ENS Lyon UCBL INRIA), 46 All\'ee d'Italie, 69364
Lyon, France}
\and
Luis Pino \institute{INRIA/DGA and LIX \'Ecole Polytechnique de Paris}
\and
Frank Valencia\institute{CNRS and LIX \'Ecole Polytechnique de Paris}}

\def\titlerunning{Reducing Weak to Strong Bisimilarity in CCP}
\def\authorrunning{A. Aristizabal, F. Bonchi, L. Pino \& F. Valencia}

\begin{document}
\maketitle

\begin{abstract}
Concurrent constraint programming (ccp) is a well-established model for
concurrency that singles out the fundamental aspects of asynchronous systems
whose agents (or processes) evolve by posting and querying (partial)
information in a global medium. Bisimilarity is a standard behavioural
equivalence in concurrency theory. However, only recently a well-behaved
notion of bisimilarity for ccp, and a ccp partition refinement algorithm
for deciding the strong version of this equivalence have been proposed.
Weak bisimiliarity is a central behavioural equivalence in process calculi
and it is obtained from the strong case by taking into account only the
actions that are observable in the system.  Typically, the standard partition
refinement can also be used for deciding weak bisimilarity simply by using
Milner's reduction from weak to strong bisimilarity; a technique referred
to as \emph{saturation}. In this paper we demonstrate  that, because of
its involved labeled transitions, the above-mentioned saturation technique
does not work for ccp. We give an alternative reduction from weak ccp
bisimilarity to the strong one that  allows us to use the ccp partition
refinement algorithm for deciding this equivalence.
\end{abstract}

\section{Introduction} \label{sec:intro}
Since the introduction of process calculi, one of the richest sources of foundational investigations stemmed from the analysis of
\emph{behavioural equivalences}:
in any formal process language, systems which are syntactically different may denote the same process, i.e., they have the same
\emph{observable behaviour}.

A major dichotomy among behavioural equivalences concerns \emph{strong} and \emph{weak} equivalences. In strong equivalences,
all the transitions performed by a system are deemed observable. In weak equivalences, instead, internal transitions
(usually denoted by $\tau$) are unobservable. On the one hand, weak equivalences are more abstract (and thus closer
to the intuitive notion of behaviour); on the other hand, strong equivalences are usually much easier to be checked
(for instance, in \cite{DBLP:journals/iandc/LanesePSS11} a strong equivalence is introduced which
is computable for a Turing complete formalism).

\emph{Strong bisimilarity} is one of the most studied behavioural equivalence and many algorithms (e.g., \cite{Victor:94:CAV,Fernandez:89:SCP,Ferrari:98:CAV})
have been developed to check whether two systems are equivalent up to strong bisimilarity.
Among these, the \emph{partition refinement algorithm}
\cite{Kanellakis:83:PODC} is one of the best known: first it generates the
state space of a labeled transition system (LTS), i.e., the set of states reachable through the transitions; then,
it creates a partition equating all states and afterwards,
iteratively, refines these partitions by splitting non equivalent
states. At the end, the resulting partition equates all and only  bisimilar states.

\emph{Weak bisimilarity} can be computed by reducing it to strong bisimilarity.
Given an LTS $\trans{\cdot}$ labeled with actions $a,b,\ldots$ one can build $\newtrans{\cdot}$ as follows.
\[\begin{array}{c}
 \bigfrac{P \trans{a}Q}{P \newtrans{a} Q} \qquad
 \bigfrac{}{P \newtrans{\tau} P} \qquad
 \bigfrac{P \newtrans{\tau} P_1 \newtrans{a} Q_1 \newtrans{\tau} Q}{P \newtrans{a} Q} \\
\end{array}\]
Since weak bisimilarity on $\tr{a}$ coincides with strong bisimilarity on $\newtrans{a}$, then one can check weak bisimilarity
with the algorithms for strong bisimilarity on the new LTS $\newtrans{a}$.

It is worth pointing out  that an alternative presentation of $\newtrans{\cdot}$ with sequences of actions as labels is also possible \cite{Milner:80:Book}. Nevertheless, the resulting transition system may be infinite-branching and hence not amenable to automatic verification using standard algorithms such as partition refinement.

\emph{Concurrent Constraint Programming} (ccp)
\cite{Saraswat:90:POPL} is a formalism that
combines the traditional algebraic and operational view of process
calculi with a declarative one based upon first-order logic. In ccp,
processes (agents or programs) interact by \emph{adding} (or \emph{telling}) and \emph{asking} information
(namely, constraints) in a medium (the \emph{store}).

Inspired by \cite{Bonchi:06:LICS,Bonchi:09:FOSSACS}, the authors introduced in \cite{Aristizabal:11:FOSSACS}
both strong and weak bisimilarity for ccp and showed that the
weak equivalence is \emph{fully abstract} with respect to the standard observational equivalence of \cite{Saraswat:91:POPL}.
Moreover, a variant of the partition refinement algorithm is given in \cite{Aristizabal:12:SAC} for checking
strong bisimilarity on (the finite fragment) of concurrent constraint programming.

In this paper, first we show that the standard method for reducing weak to strong bisimilarity does not work for ccp and then we provide
a way out of the impasse. Our solution can be readily explained by observing that the labels in the LTS of a ccp agent
are constraints (actually, they are ``the minimal constraints'' that the store should satisfy in order to make the agent progress).
These constraints form a lattice where the least upper bound (denoted by $\sqcup$) intuitively corresponds to conjunction and
the bottom element is the constraint $true$. (As expected, transitions labeled by $true$ are internal transitions,
corresponding to the $\tau$ moves in standard process calculi).
Now, rather than closing the transitions just with respect to $true$, we need to close them w.r.t. all the constraints.
Formally we build the new LTS with the following rules.
\[\begin{array}{c}
 \bigfrac{P \trans{a}Q}{P \newtrans{a} Q} \qquad
 \bigfrac{}{P \newtrans{true} P} \qquad
 \bigfrac{P \newtrans{a} Q \newtrans{b} R }{P \newtrans{a\sqcup b} R} \\
\end{array}\]
Note that, since $\sqcup$ is idempotent, if the original LTS $\trans{a}$ has finitely many transitions, then also $\newtrans{a}$ is finite.
This allows us to use the algorithm in \cite{Aristizabal:12:SAC} to check weak bisimilarity on (the finite fragment)
of concurrent constraint programming. We have implemented this procedure in a tool that is available at \url{http://www.lix.polytechnique.fr/~andresaristi/checkers/}.
To the best of our knowledge, this is the first tool for checking weak equivalence of ccp programs.

This paper is structured as follows. In Sec. \ref{sec:background} we recall the partition refinement method and the standard reduction from weak to strong bisimilarity. We also recall the ccp formalism, its equivalences, and the ccp partition refinement algorithm. We then show why the standard reduction does not work for ccp. Finally, in Sec. \ref{sec:weakCCP} we present our reduction and show its  correctness.

{\bf Related Work.} Ccp is not the only formalism where weak bisimilarity cannot be naively reduced to the strong one.
Probably the first case in literature can be found in \cite{Victor:94:CAV} that introduces
an algorithm for checking weak open bisimilarity of $\pi$-calculus. This algorithm is rather different from ours,
since it is on-the-fly \cite{Fernandez:89:SCP} and thus it checks the equivalence of only two given states (while our algorithm,
and more generally all algorithms based on partition refinement, check the equivalence of all the states of a given LTS). Also \cite{Baldan:07:TCS} defines weak labelled transitions following the above-mentioned standard method which does not work in the ccp case.

Analogous problems to the one discussed in this paper arise in Petri nets \cite{DBLP:conf/concur/Sobocinski10,DBLP:conf/concur/BruniMM11}, in tile transition systems \cite{DBLP:conf/birthday/GadducciM00,Bruni:05:CONCUR} and, more generally,
in the theory of reactive systems \cite{Jensen} (the interested reader is referred to \cite{Sobocinski12} for an overview).
In all these cases, labels form a monoid where the neutral element is the label of internal transitions.
Roughly, when reducing from weak to strong bisimilarity, one needs to close the transitions with respect to the composition of the monoid
(and not only with respect to the neutral element). However, in all these cases, labels composition is not idempotent
(as it is for ccp) and thus a finite LTS might be transformed into an infinite one. For this reason, this procedure applied
to the afore mentioned cases is not effective for automatic verification.


\section{From Weak to Strong CCP Bisimilarity: Saturation Approach } \label{sec:background}

The problem of whether two states are weakly bisimilar  in traditional labeled transitions systems
is typically reduced to the problem of whether  they are strongly bisimilar which can be
efficiently verified using partition refinement. We shall refer to this standard reduction as Milner's saturation method \cite{Aceto:11:BookChapter}.

In this section we shall show that this method does not work for ccp. More precisely,  Milner's reduction
will produce an equivalence that does not correspond to the one expected. First,
we shall recall the partition refinement algorithm for strong bisimilarity and Milner's saturation method. Then we show
the corresponding notions in ccp.



\paragraph{Standard Partition Refinement.} \label{sssec:partitionRef}
In this section we recall the partition refinement algorithm \cite{Kanellakis:83:PODC} for
checking bisimilarity over the states of a \emph{labeled transition system}.
Remember that an LTS can be intuitively seen as a graph where nodes represent states
and arcs represent transitions between states. A transition $P\tr{a}Q$ between $P$ and $Q$ labeled with $a$ can be
typically thought of as  an evolution from $P$ to $Q$ provided that a condition $a$ is met.
Transition systems can be used to represent the evolution of processes in calculi such
as CCS and the $\pi$-calculus \cite{Milner:80:Book,Milner:99:Book}.
In this case states correspond to processes and transitions are given by the operational semantics of the calculus.

Let us now introduce some notation. Given a set $S$, a \emph{partition} $\mathcal{P}$ of $S$ is a set of non-empty
\emph{blocks}, i.e., subsets of $S$, that are all disjoint and whose
union is $S$. We write $\{B_1\}\dots \{B_n\}$ to denote a partition consisting
of (non-empty) blocks $B_1, \dots, B_n$. A partition represents an equivalence relation
where equivalent elements belong to the same block. We write $P \mathcal{P} Q$ to mean that $P$ and $Q$ are equivalent in the partition
$\mathcal{P}.$

The \emph{partition refinement algorithm} (see Alg. \ref{algo:satPR}) checks bisimilarity as follows.
First, it computes $IS^{\star}$, that is the set of
all states that are reachable from the set of initial state $IS$. Then it creates the
partition $\mathcal{P}^0$ where all the elements of $IS^{\star}$ belong to
the same block (i.e., they are all equivalent). After the initialization, it iteratively refines the
partitions by employing the function $\mathbf{F}$, defined as follows:
for all partitions $\mathcal{P}$,
$P \, \mathbf{F}(\mathcal{P})\, Q$ iff
\begin{itemize}
 \item if $P\tr{a}P'$ then exists $Q'$ s.t. $Q\tr{a}Q'$ and  $P'\, \mathcal{P} Q'$.
\end{itemize}
The algorithm terminates whenever two consecutive partitions are
equivalent. In such a partition two states belong to the same block iff they are bisimilar.

%


\begin{algorithm}\caption{\texttt{Partition-Refinement($IS$)}}\label{algo:satPR}
\textbf{Initialization}
\begin{enumerate}
\item $IS^{\star}$ is the set of all processes reachable from $IS$,
\item $\mathcal{P}^0:=\{IS^{\star}\}$,
\end{enumerate}
\textbf{Iteration} $\mathcal{P}^{n+1}:=\mathbf{F}(\mathcal{P}^n)$,

\textbf{Termination} If $\mathcal{P}^n=\mathcal{P}^{n+1}$ then return $\mathcal{P}^n$.
\end{algorithm}

\paragraph{Standard reduction from weak to strong bisimilarity.} \label{sssec:weakToStrong}
As pointed out in the literature (Chapter 3 from \cite{Sangiorgi:12:Book}),
in order to compute weak bisimilarity, we can use the above mentioned partition refinement.
The idea is to start from the graph generated via the operational semantics and then
\emph{saturate} it using the rules described in Tab. \ref{tab:milnerlabsem} to produce a new
labeled transition relation $\newtrans{}$.
Recall that  $\reds$ is the reflexive and transitive closure of the transition relation $\trans{}$.
Now the problem whether two states are weakly bisimilar can be reduced to checking whether
they are strongly bisimilar wrt $\newtrans{}$ using partition refinement.
As we will show later on, this approach does not work in a formalism like concurrent constraint programming.
We shall see that the problem involves the ccp transition labels which, being constraints, can be arbitrary combined
using the lub operation $\sqcup$  to form a new one. Such a situation does
not arise in CCS-like labelled transitions.

\begin{notation}
When the label of a transition is $\true$ we will omit it.
Namely, henceforth we will use $\G \trans{} \G'$ and $\G \newtrans{} \G'$
to denote $\G \trans{\true} \G'$ and $\G \newtrans{\true} \G'$.
\end{notation}



\begin{table}[tb]
\[\begin{array}{|c|}
\hline
\makebox{MR1} \quad \bigfrac{\G \trans{\A} \G'}{\G \newtrans{\A} \G'} \qquad
\makebox{MR2} \quad \bigfrac{}{\G \newtrans{\true} \G} \qquad
\makebox{MR3} \quad \bigfrac{\G \newtrans{\true} \G_1 \newtrans{\A} \G_2 \newtrans{\true} \G'}{\G \newtrans{\A} \G'} \\
\hline
\end{array}\]
\caption{Milner's Saturation Method}
\label{tab:milnerlabsem}
\end{table}

\subsection{CCP} \label{ssec:bisimCCP}

We shall now recall ccp and the adaptation
of the partition refinement algorithm to compute bisimilarity in ccp \cite{Aristizabal:12:SAC}.

\paragraph{Constraint Systems.} The ccp model is parametric in a \emph{constraint system (cs)} specifying the
structure and interdependencies of the information that processes can ask or
and add to a \emph{central shared store}.  This information is represented
as assertions traditionally referred to as \emph{constraints}. Following \cite{Boer:95:TCS, Mendler:95:NJC} we regard a cs as a
complete algebraic lattice in which the ordering $\sqsubseteq$ is the
reverse of an entailment relation: $c\sqsubseteq d$ means $d$
\emph{entails} $c$, i.e., $d$ contains ``more information'' than $c$.  The
top element $\false$ represents inconsistency, the bottom element $\true$
is the empty constraint, and the \emph{least upper bound} (lub) $\sqcup$
is the join of information. 


\begin{definition}[cs]\label{def:constraintsystem}
A \emph{constraint system (cs)}
$ \mathbf{C} = ({\Con},{\Con}_0,\sqsubseteq,\sqcup,\true,\false)$ is a complete algebraic lattice where ${\Con},$
the set of constraints, is a partially ordered set wrt
 $\sqsubseteq$, ${\Con}_0$ is the subset of \emph{compact} elements of
${\Con}$, $\sqcup$ is the lub operation defined on all subsets, and $\true$, $\false$ are
the least and greatest elements of ${\Con}$, respectively.
\end{definition}


\begin{remark}\label{rem:well-founded} We shall assume that the constraint system is well-founded and, for practical reasons, that its ordering $\sqsubseteq$ is decidable.
\end{remark}

We now define the constraint system we  use in our examples.

\begin{example}
 Let ${\it Var}$ be a set of variables and $\omega$ be the set of natural numbers.
A variable assignment is a function $\mu: {\it Var} \longrightarrow
\omega$. We use $\mathcal{A}$ to denote the set of all
assignments, ${\mathcal P}(\mathcal{A})$ to denote the powerset of $\mathcal{A}$, $\emptyset$ the empty set and $\cap$ the intersection of sets.
Let us define the following constraint system: The set of constraints is ${\mathcal P}(\mathcal{A})$. We define $c \sqsubseteq d$  iff  $c \supseteq d$. The constraint $\false$ is $\emptyset$, while $\true$ is
$\mathcal{A}$. Given two constraints $c$ and $d$, $c \sqcup d$ is
the intersection $c\cap d$. We will often
use a formula like $x<n$ to denote the corresponding constraint, i.e.,
the set of all assignments that map $x$ to a number smaller than $n$.
\end{example}

\paragraph{Processes}

We now recall the basic ccp process constructions. For the sake of space and simplicity we dispense with
the recursion operator, which is defined in the standard way as in CCS or other process algebras, and the local/hiding operator (see \cite{Aristizabal:11:FOSSACS} for further details).

\emph{Syntax.}\label{sec:syntax} Let us presuppose a constraint system ${\bf C}=
({\Con},{\Con}_0,\sqsubseteq,\sqcup,\true,\false)$. The ccp processes are given  by the following
syntax:
\[
P,Q::= \Stop \mid \tell(c) \mid \ask(c)\rightarrow P \mid P
\parallel Q  \mid P
 +   Q
\] where $c \in {\Con}_0$. Intuitively, $\Stop$ represents termination, $\tell(c)$ adds the constraint (or partial information) $c$ to the store.
The addition is performed regardless the generation of inconsistent information.
The process $\ask(c)\rightarrow P$ may execute $P$ if $c$ is entailed from the
information in the store. The processes $P \parallel Q$ and $P \, + \,  Q$ stand, respectively, for the
{\it parallel execution} and {\it non-deterministic choice} of $P$ and $Q$.


\emph{Reduction Semantics.}\label{sec:operationalmodel} The operational semantics is given by transitions
between configurations.
 A configuration is a pair $\pairccp{P}{d}$ representing a \emph{state} of
a system;  $d$ is a constraint representing the global store, and
$P$ is a process, i.e., a term of the syntax.  We use ${\it Conf}$ with
typical elements $\gamma,\gamma',\ldots$ to denote the set of
configurations.  The operational model of ccp is given by the transition
relation $\trans{} \subseteq \Conf  \times \Conf$ defined in Tab. \ref{tab:opersem}. The rules in Tab. \ref{tab:opersem} are
easily seen to realize the above intuitions.

%
%
%
%

\begin{table}
{\scriptsize
$$\makebox{R1} \pairccp{\tellp{c}}{d} \trans{}
\pairccp{\Stop}{d \sqcup c} \quad
\makebox{R2} \bigfrac{c \sqsubseteq d}{\pairccp{\askp{(c)}{P}}{d} \trans{} \pairccp{P}{d}}
\quad
\makebox{R3} \bigfrac{\pairccp{P}{d} \trans{}
\pairccp{P'}{d'}}{\pairccp{P \parallel Q}{d} \trans{}
\pairccp{P'\parallel Q}{d'}} \quad
\makebox{R4} \bigfrac{\pairccp{P}{d} \trans{}
\pairccp{P'}{d'}}{\pairccp{P \, + \, Q}{d} \trans{}
\pairccp{P'}{d'}}
$$

}
\caption{Reduction semantics for ccp (the symmetric rules for R3 and R4 are omitted).} \label{tab:opersem}
\end{table}

\emph{Barbed Semantics.}\label{sec:barbsem}
The authors in \cite{Aristizabal:11:FOSSACS} introduced a  barbed semantics for ccp.
Barbed equivalences have been introduced in \cite{Milner:92:ICALP}
for CCS, and  have become the standard behavioural equivalences for
formalisms equipped with unlabeled reduction semantics. Intuitively,
\emph{barbs} are basic observations (predicates) on the states of a
system. In the case of ccp, barbs are taken from the underlying set $\Con_0$ of the constraint system.
A configuration $\gamma=\pairccp{P}{d}$ is said to \emph{satisfy} the barb $c$
($\gamma \downarrow_c$) iff $c \sqsubseteq d$. Similarly, $\G$ satisfies
a \emph{weak barb} $c$ ($\G\wbarb{c})$ iff there exist $\G'$ s.t. $\G \reds \G'\barb{c}$.

In this context, the equivalence proposed is the
\emph{saturated bisimilarity} \cite{Bonchi:06:LICS, Bonchi:09:FOSSACS}. 
Intuitively, in order for two states to be saturated bisimilar, then (i) they should expose the same barbs,
(ii) whenever one of them moves then the other should reply and arrive at an equivalent state (i.e. follow the bisimulation game),
(iii) they should be equivalent under all the possible contexts of the language.
In the case of ccp, it is enough to require that bisimulations are \emph{upward closed} as in condition $(iii)$ below.

\begin{definition}[Saturated Barbed Bisimilarity]
\label{def:satbarbbis}
A saturated barbed bisimulation is a symmetric relation
$\R$ on configurations s.t. whenever
$(\gamma_1, \gamma_2) \in \R$  with $\gamma_1=\conf{P}{c}$ and $\gamma_2=\conf{Q}{d}$ implies that:
$(i)$ if $\gamma_1 \downarrow_e$ then $\gamma_2 \downarrow_e$,
$(ii)$ if $\gamma_1 \trans{} \gamma_1'$ then there exists
$\gamma_2'$ s.t. $\gamma_2 \trans{} \gamma_2'$ and $(\gamma_1',
\gamma_2') \in \R$,
$(iii)$ for every $a\in \Con_0$, $(\pairccp{P}{c \sqcup a},
\pairccp{Q}{d \sqcup a}) \in \mathcal{R}$.
We say that  $\gamma_1$ and $\gamma_2$ are saturated barbed bisimilar ($\gamma_1  \;
\satbis \; \gamma_2$) if there exists a saturated barbed bisimulation $\R$ s.t.
$(\gamma_1,\gamma_2) \in \mathcal{R}$.
\end{definition}

We use the term ``saturated'' to be consistent with the original idea in \cite{Bonchi:06:LICS, Bonchi:09:FOSSACS}.
However, ``saturated'' in this context has nothing to do with the Milner's ``saturation'' for weak bisimilarity.
In the following, we will continue to use ``saturated'' and ``saturation'' to denote these two different concepts.

\begin{example} \label{ex:satbarbbis1}
Take $T=\tellp{\true}$, $P=\askp{(x<7)}{T}$ and $Q=\askp{(x<5)}{T}$. You can see that
$\pairccp{P}{true} \not \!\! \satbis \pairccp{Q}{true}$, since $\pairccp{P}{x<7} \tr{}$, while
$\pairccp{Q}{x<7} \not \! \! \tr{}$. Consider now the configuration $\pairccp{P+Q}{true}$ and
observe that $\pairccp{P+Q}{true} \satbis \pairccp{P}{true}$. Indeed, for all constraints $e$, s.t.
$x<7 \sqsubseteq e$, both the configurations evolve into $\pairccp{T}{e}$, while for all $e$ s.t.
$x<7 \not \sqsubseteq e$, both configurations cannot proceed. Since $x<7 \sqsubseteq x<5$,
the behaviour of $Q$ is somehow absorbed by the behaviour of $P$.
\end{example}

As we mentioned before, we are interested in deciding the weak version of the notion above.
Then, \emph{weak saturated barbed bisimilarity} ($\wsatbis$) is obtained from
Def. \ref{def:satbarbbis} by replacing the strong barbs
in condition $(i)$ for its weak version ($\wbarb{}$) and the transitions in condition $(ii)$
for the reflexive and transitive closure of the transition relation ($\reds$).




\begin{table}[t!]
{\scriptsize
$$
\makebox{LR1}\pairccp{\tellp{c}}{d} \trans{\true}
\pairccp{\Stop}{d \sqcup c} \quad
\makebox{LR2}\bigfrac{\alpha \in \min \{a\in \Con_0 \, | \, c
\sqsubseteq d \sqcup a \ \}} {\pairccp{\askp{(c)}{P}}{d}
\trans{\alpha} \pairccp{P}{d \sqcup \alpha}} \quad
\makebox{LR3}
\bigfrac{\pairccp{P}{d} \trans{\alpha} \pairccp{P'}{d'}}
{\pairccp{P\parallel Q}{d} \trans{\alpha} \pairccp{P'\parallel
Q}{d'}} \quad
\makebox{LR4}
\bigfrac{\pairccp{P}{d} \trans{\alpha} \pairccp{P'}{d'}}
{\pairccp{P + Q}{d} \trans{\alpha} \pairccp{P'}{d'}}
$$

}
\caption{Labeled semantics for ccp (the symmetric rules for LR3 and LR4 are omitted).}\label{tab:labsem}
\end{table}

\begin{figure}
\begin{center}
{\small
\begin{tikzpicture}
\node (defT) at (-1.1,4.2) {{\footnotesize $T = \tellp{\truep}$}};
\node (defT') at (-1,3.8) {{\footnotesize $T' = \tellp{y=1}$}};
\node (defP) at (2,4.2) {{\footnotesize $P = \askp{(x<7)}{T}$}};
\node (defS) at (2,3.8) {{\footnotesize $S = \askp{(z<7)}{P}$}};
\node (defQ) at (5.5,4.2) {{\footnotesize $Q = \askp{(x<5)}{T}$}};
\node (defQ') at (5.6,3.8) {{\footnotesize $Q' = \askp{(x<5)}{T'}$}};
\node (defR) at (9.5,4.2) {{\footnotesize $R = \askp{(z<5)}{(P+Q)}$}};
\node (defR') at (9.6,3.8) {{\footnotesize $R' = \askp{(z<5)}{(P+Q')}$}};
\node (RS) at (-1,1) {$\pairccp{R+S}{\truep}$};
\node (S) at (-1,2) {$\pairccp{S}{\truep}$};
\node (R'S) at (-1,3) {$\pairccp{R'+S}{\truep}$};
\node (PQ') at (3,3) {$\pairccp{P+Q'}{z<5}$};
\node (P) at (3,2) {$\pairccp{P}{z<7}$};
\node (PQ) at (3,1) {$\pairccp{P+Q}{z<5}$};
\node (P2) at (3,0) {$\pairccp{P}{z<5}$};
\node (T') at (6.5,3) {$\pairccp{T'}{z<5 \sqcup x<5}$};
\node (T1) at (6.5,2) {$\pairccp{T}{z<7 \sqcup x<7}$};
\node (T2) at (6.5,1) {$\pairccp{T}{z<5 \sqcup x<5}$};
\node (T3) at (6.5,0) {$\pairccp{T}{z<5 \sqcup x<7}$};
\node (S1) at (11,3) {$\pairccp{\stopp}{z<5 \sqcup x<5 \sqcup y=1}$};
\node (S2) at (11,2) {$\pairccp{\stopp}{z<7 \sqcup x<7}$};
\node (S3) at (11,1) {$\pairccp{\stopp}{z<5 \sqcup x<5}$};
\node (S4) at (11,0) {$\pairccp{\stopp}{z<5 \sqcup x<7}$};
\draw[->] (PQ') to [out=90, in=0] (2.5,3.5) -- (-2.2,3.5) -- (-2.2,-0.5) -- node[above] {{\scriptsize $x<7$}} (6,-0.5) to [out=0,in=270] (T3);
\draw[->] (R'S) to node[above] {{\scriptsize $z<5$}} (PQ');
\draw[->] (R'S) to node[above] {{\scriptsize $z<7$}} (P);
\draw[->] (S) to node[above] {{\scriptsize $z<7$}} (P);
\draw[->] (RS) to node[above] {{\scriptsize $z<5$}} (PQ);
\draw[->] (RS) to node[above] {{\scriptsize $z<7$}} (P);
\draw[->] (PQ') to node[above] {{\scriptsize $x<5$}} (T');
\draw[->] (P) to node[above] {{\scriptsize $x<7$}} (T1);
\draw[->] (PQ) to node[above] {{\scriptsize $x<5$}} (T2);
\draw[->] (PQ) to node[above] {{\scriptsize $x<7$}} (T3);
\draw[->] (P2) to node[above] {{\scriptsize $x<7$}} (T3);
\draw[->] (T') to node[above] {{\scriptsize $\truep$}} (S1);
\draw[->] (T1) to node[above] {{\scriptsize $\truep$}} (S2);
\draw[->] (T2) to node[above] {{\scriptsize $\truep$}} (S3);
\draw[->] (T3) to node[above] {{\scriptsize $\truep$}} (S4);
\end{tikzpicture}}
\end{center}
\caption{The LTS of the running example ($IS = \{\pairccp{R'+S}{\true},\pairccp{S}{\true},\pairccp{R+S}{\true}\}$).}
\label{fig:ltsexample}
\end{figure}
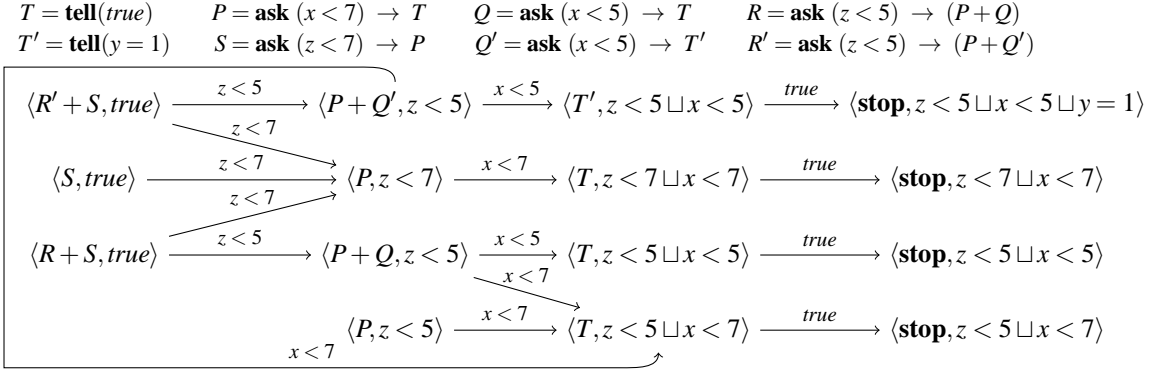

\emph{Labeled Semantics.} As explained in \cite{Aristizabal:11:FOSSACS}, in a transition of the form
$\pairccp{P}{d} \trans{\alpha} \pairccp{P'}{d'}$ the label $\alpha$
represents a \emph{minimal} information (from the environment)
that needs to be added to the store $d$ to evolve from  $\pairccp{P}{d}$ into $\pairccp{P'}{d'}$, i.e.,
$\pairccp{P}{d \sqcup \alpha} \trans{} \pairccp{P'}{d'}$.
The labeled transition relation $\rrarrow\;\subseteq{\it Conf}\times \Con_0
\times{\it Conf}$ is defined by the rules in Tab. \ref{tab:labsem}.
The rule LR2, for example, says that $\pairccp{\askp{(c)}{P}}{d}$ can evolve to
$\pairccp{P}{d\sqcup \alpha}$ if the environment provides a minimal
constraint $\alpha$ that added to the store $d$ entails $c$, i.e.,
$\alpha \in \min \{a \in \Con_0 \, | \, c \sqsubseteq d \sqcup a
\}$. Note that assuming that $(\Con, \sqsubseteq)$ is well-founded
(Remark \ref{rem:well-founded}) is necessary to guarantee that $\alpha$ exists
whenever $ \{a \in \Con_0 \, | \, c \sqsubseteq d \sqcup a \ \}$ is
not empty.  The other rules are easily seen to realize the above intuition.
Fig. \ref{fig:ltsexample} illustrates the LTSs of our running example.

The labeled semantics is \emph{sound} and \emph{complete} wrt the
unlabeled one. Soundness states that $\pairccp{P}{d}
\trans{\alpha}\pairccp{P'}{d'}$ corresponds to our intuition that if
$\alpha$ is added to $d$, $P$ can reach $\pairccp{P'}{d'}$.
Completeness states that if we add $a$ to (the store in)
$\pairccp{P}{d}$ and reduce to $\pairccp{P'}{d'}$, it exists a
minimal information $\alpha \sqsubseteq a$ such that $\pairccp{P}{d}
\trans{\alpha} \pairccp{P'}{d''}$ with $d'' \sqsubseteq d'$.


The following lemma is an extension of the one in \cite{Aristizabal:11:FOSSACS}
which considers nondeterministic ccp. 

\begin{lemma}[Correctness of $\trans{}$]
\label{lem:correctness}
(Soundness) If $\transition{P}{c}{\A}{P'}{c'}$ then $\transition{P}{c \lub \A}{}{P'}{c'}$.
(Completeness) If $\transition{P}{c \lub a}{}{P'}{c'}$ then there exists $\A$ and $b$ s.t. $\transition{P}
{c}{\A}{P'}{c''}$ where $\A \lub b = a$ and $c'' \lub b = c'$.
\end{lemma}




The above lemma  is central for deciding bisimilarity
in ccp. In fact, we will show later that for the weak (saturated) semantics the completeness direction
does not hold. From this we will show that the standard reduction from weak to strong does not work.

\subsubsection{Equivalences: Saturated Barbed, Irredundant and Symbolic Bisimilarity}
In this section we recall how to check $\satstbisim$ with a modified version of partition refinement introduced in \cite{Aristizabal:12:SAC}. Henceforth, we shall refer to this version as \emph{ccp partition refinement (ccp-PR)}.

The main problem with  checking $\satstbisim$ is the quantification over all contexts. This problem is addressed in \cite{Aristizabal:12:SAC}
following the abstract approach in \cite{Bonchi:09:ESOP}. More precisely,
we use an equivalent notion, namely \emph{irredundant bisimilarity} $\irrbis$,  which
can be verified  with ccp-PR. As its name suggests,
$\irrbis$  only takes into account those transitions deemed irredundant.\footnote{Redundancy
itself is not trivial to check, for more information go to \cite{Aristizabal:12:SAC}.}
However, technically speaking, going from $\satstbisim$ to $\irrbis$ requires one
intermediate notion, so-called \emph{symbolic bisimilarity}.  These three notions
are shown to be equivalent, i.e.,   \( \satstbisim = \symbis = \irrbis.\)  In the following we recall all of them.

Let us first give some auxiliary definitions. The first concept is that of \emph{derivation}.
Consider the following transitions (taken from Fig. \ref{fig:ltsexample}):
\[\mbox{(a) } \pairccp{P+Q}{z<5} \trans{x<7} \pairccp{T}{z<5 \lub x<7} ~~~~~~~~~~~~
\mbox{(b) } \pairccp{P+Q}{z<5} \trans{x<5} \pairccp{T}{z<5 \lub x<5} \]
Transition (a) means that for all constraints $e$ s.t. $x<7$ is entailed by $e$ (formally $x<7 \entailed e$),
the transition (c) $\pairccp{P+Q}{z<5 \lub e} \trans{} \pairccp{T}{z<5 \lub e}$ can be performed,
while transition (b) means that the reduction (c) is possible for all $e$ s.t. $x<5 \entailed e$.
Since $x<7 \entailed x<5$, transition (b) is ``redundant'', in the sense that its meaning is ``logically derived'' by transition (a).
The following notion captures the above intuition:
\begin{definition}[Derivation $\deriv$]
\label{def:deriv}
We say that the transition $t = \transition{P}{c}{\A}{P'}{c'}$ derives $t' =
\transition{P}{c}{\B}{P'}{c''}$ (written $t \deriv t'$) iff there exists $e$ s.t. $\A \lub e = \B$ and $c' \lub e = c''$.
\end{definition}

One can verify in the above example that (a) $\deriv$ (b), and
notice that both transitions arrive at the same process $P'$,
the difference lies in the label and the store.
Now imagine the situation where the initial configuration is able
to perform another transition with $\B$ (as in $t'$),
let us also assume that such transition arrives at a configuration which is
equivalent to the result of $t'$. Therefore,
it is natural to think that, since $t$ dominates $t'$,
such new transition should also be dominated by $t$.
Let us explain with an example, consider the two following transitions:
\[\mbox{(e) } \pairccp{R+S}{\true} \trans{z<7} \pairccp{P}{z<7} ~~~~~~~~~~~~
\mbox{ (f) }\pairccp{R+S}{\true} \trans{z<5} \pairccp{P+Q}{z<5} \]
Note that transition (f) cannot be derived by other transitions, since (e) $\not \deriv$ (f).
Indeed, $P$ is syntactically different from $P+Q$, even if they have the same behaviour when inserted in the store $z<5$, i.e.,
$\pairccp{P}{z<5}\satbis \pairccp{P+Q}{z<5}$ (since $\satbis$ is upward closed). Transition (f)
is also ``redundant'', since its behaviour ``does not add anything'' to the behavior of (e).
The following definition encompasses this situation:

\begin{definition}[Derivation w.r.t $\R$, $\derivR$]
\label{def:derivR}
We say that the transition $t = \G \trans{\A} \G_1$ derives $t' =
\G \trans{\B} \G_2$ w.r.t. to $\R$ (written $t \derivR t'$) iff
there exists $\G'_2$ s.t. $t \deriv \G \trans{\B} \G'_2$ and $\G'_2 \R \G_2$.
\end{definition}

Then, when $\R$ represents some sort of equivalence, this notion will
capture the situation above mentioned.
Notice that $\deriv$ is $\derivR$ with $\R$ being the identity relation ($\idrel$).
Now we introduce the concept of \emph{domination}, which consists in strengthening the
notion of derivation by requiring  labels to be different.

\begin{definition}[Domination $\dom$]
\label{def:dom}
We say that the transition $t = \transition{P}{c}{\A}{P'}{c'}$ dominates $t' = \transition{P}{c}{\B}{P'}{c''}$ (written $t \dom t'$) iff $t \deriv t'$ and $\A \neq \B$.
\end{definition}

Similarly, as we did for derivation, we can define domination
depending on a relation. Again, $\dom$ is just $\domR$ when
$\R$ is the identity relation ($\idrel$).

\begin{definition}[Redundancy and Domination w.r.t $\R$, $\domR$]
\label{def:domR}
We say that the transition $t = \transition{P}{c}{\A}{P'}{c'}$ dominates $t' = \transition{P}{c}{\B}{Q}{d}$ w.r.t. to $\R$ (written $t \domR t'$) iff there exists $c''$ s.t. $t \dom \transition{P}{c}{\B}{P'}{c''}$ and $\conf{P}{c''} \R \conf{Q}{d}$. Also, a transition is said to be redundant when it is dominated
by another, otherwise it is said to be irredundant.
\end{definition}

We are now able to introduce symbolic bisimilarity.
Intuitively, two configurations $\G_1$ and $\G_2$
are symbolic bisimilar iff (i) they have the same barbs and
(ii) whenever there is a transition from $\G_1$ to $\G'_1$ using $\A$, then
we require that $\G_2$ must reply with a similar transition $\G_2 \trans{\A} \G'_2$
(where $\G'_1$ and $\G'_2$ are now equivalent) or some other transition that derives it.
In other words, the move from the defender does not need to use exactly the same label,
but a transition that is ``stronger'' (in terms of derivation $\deriv$) could also do the job.
Formally we have the definition below.

\begin{definition}[Symbolic Bisimilarity]
\label{def:symbis}
A symbolic bisimulation is a symmetric relation
$\R$ on configurations s.t. whenever
$(\gamma_1, \gamma_2) \in \R$  with $\gamma_1=\conf{P}{c}$ and $\gamma_2=\conf{Q}{d}$ implies that:
$(i)$ if $\gamma_1 \downarrow_e$ then $\gamma_2 \downarrow_e$,
$(ii)$ if $\conf{P}{c} \trans{\A} \conf{P'}{c'}$ then there exists a transition $t = \transition{Q}{d}{\B}{Q'}{d''}$ and a store $d'$ s.t. $t \deriv \transition{Q}{d}{\A}{Q'}{d'}$ and $\conf{P'}{c'} \R \conf{Q'}{d'}$
We say that $\gamma_1$ and $\gamma_2$ are symbolic bisimilar ($\gamma_1  \; \symbis \; \gamma_2$) if there exists a symbolic bisimulation $\R$ s.t. $(\gamma_1,\gamma_2) \in \mathcal{R}$.
\end{definition}


\begin{example}\label{ex:symbis1}
\emph{To illustrate the notion of $\symbis$ we take $\pairccp{P+Q}{true}$ and $\pairccp{P}{true}$ from  Ex. \ref{ex:satbarbbis1}. We provide a symbolic bisimulation $\mathcal{R} = \{(\pairccp{P + Q}{\true},\pairccp{P}{\true})\} \cup id$ to prove $\pairccp{P+Q}{true} \symbis \pairccp{P}{true}.$  We take the pair $(\pairccp{P+Q}{true},\pairccp{P}{true})$. The first condition in Def. \ref{def:symbis} is trivial. For the second one, we take $\pairccp{P + Q}{\true} \trans{x<5} \pairccp{T}{x<5}$ and one can find transitions $t = \pairccp{P}{\true} \trans{x<7} \pairccp{T}{x<7}$ and $t' = \pairccp{P}{\true} \trans{x<5} \pairccp{T}{x<5}$ s.t. $t \deriv t'$ and $\conf{T}{x<5} \R \conf{T}{x<5}$. The restant pairs are trivially verified.}
\end{example}

And finally, the irredundant version, which follows the standard bisimulation
game where labels need to be matched, however only those transitions
so-called irredundant must be considered.

\begin{definition}[Irredundant Bisimilarity]
\label{def:irrbis}
An irredundant bisimulation is a symmetric relation
$\R$ on configurations s.t. whenever
$(\gamma_1, \gamma_2) \in \R$ implies that:
$(i)$ if $\gamma_1 \downarrow_e$ then $\gamma_2 \downarrow_e$,
$(ii)$ if $\gamma_1 \trans{\A} \gamma'_1$ and it is irredundant in $\R$ then there exists $\gamma'_2$ s.t. $\gamma_2 \trans{\A} \gamma'_2$ and $(\gamma'_1, \gamma'_2) \in \R$.
We say that $\gamma_1$ and $\gamma_2$ are irredundant bisimilar ($\gamma_1 \; \irrbis \; \gamma_2$) if there exists an irredundant bisimulation $\R$ s.t.
$(\gamma_1,\gamma_2) \in \mathcal{R}$.
\end{definition}

\begin{example}\label{ex:irrbis1}
\emph{We can verify that the relation $\mathcal{R}$ in Ex. \ref{ex:symbis1} is an irredundant bisimulation to show that $\pairccp{P+Q}{true} \irrbis \pairccp{P}{true}.$  We take the pair $(\pairccp{P + Q}{\true},\pairccp{P}{\true})$. The first item in Def. \ref{def:irrbis} is obvious. Then take $\pairccp{P + Q}{\true} \trans{x<7} \pairccp{T}{x<7}$, which is irredundant according to Def. \ref{def:domR}, then there exists a $\pairccp{T}{x<7}$ s.t. $\pairccp{P}{\true} \trans{x<7} \pairccp{T}{x<7}$ and $(\pairccp{T}{x<7},\pairccp{T}{x<7}) \in \mathcal{R}$. The other pairs are trivially proven. Notice that $\pairccp{P + Q}{\true} \trans{x<7} \pairccp{T}{x<7} \domR \pairccp{P + Q}{\true} \trans{x<5} \pairccp{T}{x<5}$ hence $\pairccp{P + Q}{\true} \trans{x<5} \pairccp{T}{x<5}$ is redundant, thus it does not need to be matched by $\conf{P}{\true}$.}
\end{example}

As we said at the beginning, the above-defined equivalences coincide with $\satbis$. The proof, given in \cite{Aristizabal:12:SAC},
strongly relies on Lemma \ref{lem:correctness}.
\begin{theorem}
$\conf{P}{c}  \irrbis \conf{Q}{d}$  iff $\conf{P}{c} \symbis \conf{Q}{d}$ iff $\conf{P}{c} \satbis \conf{Q}{d}$
\end{theorem}

\subsubsection{Partition Refinement for CCP} \label{sssec:partRefCCP}

In \cite{Aristizabal:12:SAC} the authors introduced an algorithm for checking $\satstbisim$,
by modifying the partition refinement algorithm so that to exploit $\irrbis$.
First, since configurations satisfying different barbs are
surely different, it can be safely started with a partition that equates all and only those states
satisfying the same barbs. Note that two configurations satisfy the same barbs iff they have the same store.
Thus, we take as initial partition
$\mathcal{P}^0=\{IS^{\star}_{d_1}\}\dots \{IS^{\star}_{d_n}\}$, where $IS^{\star}_{d_i}$ is the subset of
the configurations of $IS^{\star}$ with store $d_i$.\footnote{In fact,
in order to check redundancy, some new states should be added to the initial ones (hence the subscript $new$ in $IS_{new}^{\star}$).
The details of the computation are omitted given that they are not relevant for this paper,
however the interested reader is referred to \cite{Aristizabal:12:SAC} for more information.}
Secondly, instead of using the function $\mathbf{F}$ of Alg. \ref{algo:satPR},
the partitions are refined by employing the function $\mathbf{IR}$ defined as follows:
for all partitions $\mathcal{P}$,
$\gamma_1 \, \mathbf{IR}(\mathcal{P})\, \gamma_2$ iff
\begin{itemize}
 \item if $\gamma_1\tr{\alpha}\gamma_1'$ is irredundant in $\mathcal{P}$, then there exists $\gamma_2'$ s.t. $\gamma_2\tr{\alpha}\gamma_2'$ and  $\gamma_1'\, \mathcal{P} \gamma_2'$.
\end{itemize}
These two steps are the main idea behind the computation of $\irrbis$ (Alg. \ref{algo:ccp}).

%

\begin{algorithm}\caption{\texttt{CCP-Partition-Refinement($IS$)}}\label{algo:ccp}
\textbf{Initialization}
\begin{enumerate}
\item Compute $IS_{new}^{\star}$ 
\item $\mathcal{P}^0:=\{IS^{\star}_{d_1}\}\dots \{IS^{\star}_{d_n}\}$,
\end{enumerate}
\textbf{Iteration} $\mathcal{P}^{n+1}:=\mathbf{IR}(\mathcal{P}^n)$

\textbf{Termination} If $\mathcal{P}^n=\mathcal{P}^{n+1}$ then return $\mathcal{P}^n$.
\end{algorithm}

\subsection{Incompleteness of Milner's saturation method in ccp} \label{ssec:CExMilner}

As mentioned at the beginning of this section, the standard approach for
deciding weak equivalences is to add some transitions to the original processes,
so-called \emph{saturation}, and then check for the strong equivalence.
In calculi like CCS, such saturation consists in forgetting
about the internal actions that make part of a sequence containing
one observable action (Tab. \ref{tab:milnerlabsem}).
However, for ccp this method does not work.
The problem is that the transition relation proposed by Milner is not complete
for ccp, hence the relation among the saturated, symbolic and irredundant equivalences
is broken. In the next section we will provide a stronger saturation, which
is complete, and allow us to use the ccp-PR to compute $\wsatbis$.

Let us show why Milner's approach does not work.
First, we need to introduce formally the concept of \emph{completeness}
for a given transition relation.

\begin{definition}
\label{def:completeness}
We say that a transition relation $\gentrans{} \subseteq \Conf \times \Con_0 \times \Conf$ is complete
iff whenever $\conf{P}{c \lub a} \gentrans{} \conf{P'}{c'}$ then there exist $\A, b \in \Con_0$ s.t.
$\conf{P}{c} \gentrans{\A} \conf{P'}{c''}$ where $\A \lub b = a$ and $c'' \lub b = c'$.
\end{definition}

Notice that $\trans{}$ (i.e the reduction semantics, see Table \ref{tab:opersem})
is complete, and it corresponds to the second item of Lemma \ref{lem:correctness}.
Now Milner's method defines a new transition relation  $\newtrans{}$ using the rules in
Tab. \ref{tab:milnerlabsem}, but it turns out not to be complete.

\begin{figure}
\begin{tikzpicture}
[inner sep=1mm]
\node (A) at (0,0) {$\conf{\askp{\A}{(\askp{\B}{\Stop})}}{\A \lub \B}$};
\node (B) at (0,-1) {$\conf{\askp{\B}{\Stop}}{\A \lub \B}$};
\node (C) at (0,-2) {$\conf{\Stop}{\A \lub \B}$};
\node (A') at (8,0) {$\conf{\askp{\A}{(\askp{\B}{\Stop}})}{\true}$};
\node (B') at (8,-1) {$\conf{\askp{\B}{\Stop}}{\A}$};
\node (C') at (8,-2) {$\conf{\Stop}{\A \lub \B}$};
\draw[->] (A) to (B);
\draw[->] (B) to (C);
\draw[->, out=180, in=225, bend right, dotted] (A) to (-2,-1.5) -- (C);
\draw[->] (A') to node[right] {{\scriptsize $\A$}} (B');
\draw[->] (B') to node[right] {{\scriptsize $\B$}} (C');
\draw[->, out=180, in=225, bend right, dashed] (A') to node[left] {{\scriptsize missing}} (6,-1.5) -- node[above] {{\scriptsize $\A \lub \B$}} (C');
\end{tikzpicture}
\caption{Counterexample for completeness using Milner's saturation method (cycles from MR2 omitted).
Both graphs are obtained by applying the rules in Tab. \ref{tab:milnerlabsem}.}
\label{fig:milnerCompleteness}
\end{figure}
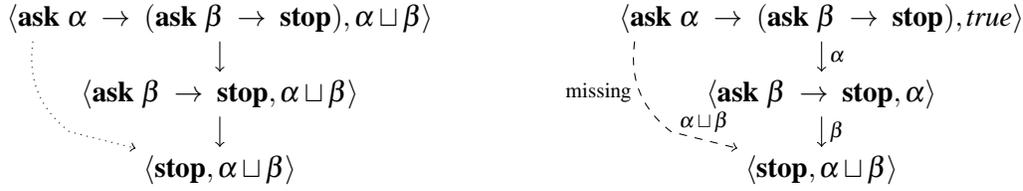

\begin{proposition}
\label{prop:milnerComp}
The relation $\newtrans{}$ defined in Table \ref{tab:milnerlabsem} is not complete.
\begin{proof}
We will show a counter-example where the completeness for $\newtrans{}$ does not hold.
Let $P=\askp{\A}{(\askp{\B}{\Stop})}$ and $d = \A \lub \B$. Now consider the transition
$\newtransition{P}{d}{}{\stopp}{d}$ and let us apply the
completeness lemma, we can take $c = \true$ and $a = \A \lub \B$, therefore
by completeness there must exist $b$ and $\C$ s.t. $\newtransition{P}{\true}{\C}{\stopp}{c''}$
where $\C \lub b = \A \lub \B$ and $c'' \lub b = d$. However, notice that the only transition
possible is $\newtransition{P}{\true}{\A}{\askp{\B}{\Stop}}{\A}$, hence completeness
does not hold since there is no transition from $\conf{P}{\true}$ to $\conf{\stopp}{c''}$ for some $c''$.
Fig. \ref{fig:milnerCompleteness} illustrates the problem.
\end{proof}
\end{proposition}

We can now use this fact to see why the method does not work
for computing $\wsatbis$ using ccp-PR. First, let us redefine some
concepts using the new transition relation $\newtrans{}$. Because of condition
(i) in $\wsatbis$, we need a new definition of barbs,
namely \emph{weak barbs w.r.t. $\newtrans{}$}.

\begin{definition}
We say $\G$ has a weak barb $e$ w.r.t. $\newtrans{}$
(written $\G \genbarb{e}$) iff $\G \newreds \G' \barb{e}$.
\end{definition}


Using this notion, we introduce Symbolic and Irredundant bisimilarity
w.r.t. $\newtrans{}$, denoted by $\newsymbis$ and $\newirrbis$ respectively.
They are defined as in Def. \ref{def:symbis} and \ref{def:irrbis} where in
condition (i) weak barbs ($\wbarb{}$) are replaced with $\genbarb{}$ and in
condition (ii) the transition relation is now $\newtrans{}$.

One would expect that since $\satstbisim = \symbis = \irrbis$ then
the natural consequence will be that $\wsatbis = \newsymbis = \newirrbis$,
given that these new notions are supposed to be the weak versions of the former ones
when using the saturation method.
However, completeness is necessary for proving $\satstbisim = \symbis = \irrbis$,
and from Proposition \ref{prop:milnerComp} we know that $\newtrans{}$ is not complete
hence we might expect $\wsatbis \neq \newsymbis \neq \newirrbis$.
In fact, the following counter-example shows these inequalities.

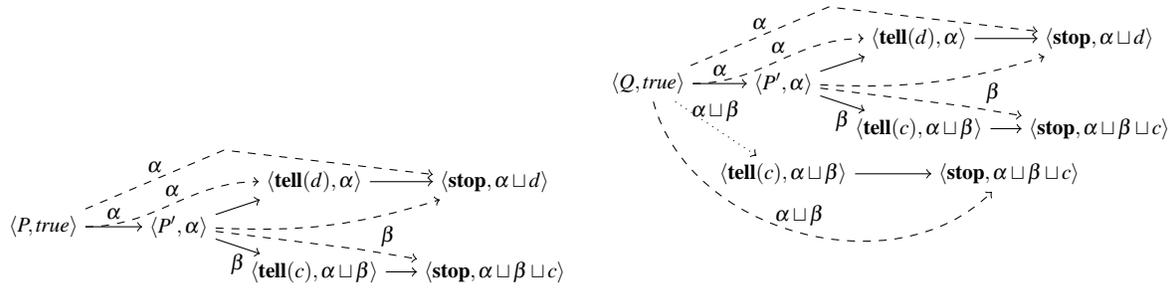
\begin{figure}[tb]
\begin{center}
\begin{tabular}{ll}
\begin{tikzpicture}[scale=0.6]
\scriptsize
\node (A) at (0, 0) {$\pairccp{P}{\true}$};
\node (B) at (3, 0) {$\pairccp{P'}{\alpha}$};
\node (C) at (6,1) {$\pairccp{\tellp{d}}{\alpha}$};
\node (D) at (6,-1) {$\pairccp{\tellp{c}}{\alpha \sqcup \beta}$};
\node (E) at (10, -1) {$\pairccp{\stopp}{\alpha \sqcup \beta \sqcup  c}$};
\node (F) at (10, 1) {$\pairccp{\stopp}{\alpha \sqcup d}$};
\draw[->] (A) to node[above] {$\alpha$} (B);
\draw[->] (B) to  (C);
\draw[->] (B) to node[below] {$\beta$} (D);
\draw[->] (C) to  (F);
\draw[->] (D) to  (E);
\draw[->,dashed] (A) to [out=0, in=180] node[above]{{$\alpha$}} (C);
\draw[->,dashed] (A) to node[above]{{$\alpha$}} (4,1.7) -- (F);
\draw[->,dashed] (B) to[bend right=10] (F);
\draw[->,dashed] (B) to (7,-0.5) -- node[above] {{$\beta$}} (E);
\end{tikzpicture} &
\begin{tikzpicture}[scale=0.6]
\scriptsize
\node (A) at (0, 0) {$\pairccp{Q}{\true}$};
\node (B) at (3, 0) {$\pairccp{P'}{\alpha}$};
\node (C) at (6,1) {$\pairccp{\tellp{d}}{\alpha}$};
\node (D) at (6,-1) {$\pairccp{\tellp{c}}{\alpha \sqcup \beta}$};
\node (E) at (10, -1) {$\pairccp{\stopp}{\alpha \sqcup \beta \sqcup  c}$};
\node (F) at (10, 1) {$\pairccp{\stopp}{\alpha \sqcup d}$};
\node (G) at (3, -2) {$\pairccp{\tellp{c}}{\alpha \sqcup \beta}$};
\node (H) at (8, -2) {$\pairccp{\stopp}{\alpha \sqcup \beta \sqcup c}$};
\draw[->] (A) to node[above] {$\alpha$} (B);
\draw[->] (B) to  (C);
\draw[->] (B) to node[below] {$\beta$} (D);
\draw[->] (C) to  (F);
\draw[->] (D) to  (E);
\draw[->,dotted] (A) to node[above] {$\alpha \sqcup \beta$} (G);
\draw[->] (G) to  (H);
\draw[->,dashed] (A) to[bend right=60] node[above]{{$\alpha \sqcup \beta$}} (H);
\draw[->,dashed] (A) to [out=0, in=180] node[above]{{$\alpha$}} (C);
\draw[->,dashed] (A) to node[above]{{$\alpha$}} (4,1.7) -- (F);
\draw[->,dashed] (B) to[bend right=10] (F);
\draw[->,dashed] (B) to (7,-0.5) -- node[above] {{$\beta$}} (E);
\end{tikzpicture}
\end{tabular}
\caption{Let $P = \askp{(\alpha)}{P'}$, $P' = (\askp{(\beta)}{\tellp{c}}) + (\askp{(\true)}{\tellp{d}})$ and $Q = P +(\askp{(\alpha \sqcup \beta)}{\tellp{c}})$.
The figure represents $\conf{P}{\true}$ and $\conf{Q}{\true}$ after being saturated using Milner's method (cycles from MR2 ommited).
The dashed transitions are the new ones added by the rules in Tab. \ref{tab:milnerlabsem}.
The dotted transition is the (irredundant) one that $\conf{Q}{\true}$ can take but $\conf{P}{\true}$ cannot
match, therefore showing that $\conf{P}{\true} \nonewirrbis \conf{Q}{\true}$}
\label{fig:milnerSat}
\end{center}
\end{figure}

\begin{example}
Let $P, P'$ and $Q$ as in Fig. \ref{fig:milnerSat}.
The figure shows $\conf{P}{\true}$ and $\conf{Q}{\true}$ after we saturate them using Milner's method.
First, notice that $\conf{P}{\true} \wsatbis \; \conf{Q}{\true}$,
since there exists a saturated weak barbed bisimulation {\small$\mathcal{R} =\{(\pairccp{P}
{\true},\pairccp{Q}{\true})\} \cup id$}.
However, $\conf{P}{\true} \nonewirrbis \conf{Q}{\true}$.
To prove that, we need to pick an irredundant transition from $\conf{P}{\true}$ or
$\conf{Q}{\true}$ (after saturation) s.t. the other cannot match.
Thus, take $\pairccp{Q}{\true} \trans{\alpha \sqcup \beta} \pairccp{\tellp{c}}{\alpha \sqcup \beta}$
which is irredundant and given that $\conf{P}{\true}$ does not have a transition with $\alpha \sqcup \beta$
then we know that there is no irredundant bisimulation containing $(\conf{P}{\true}, \conf{Q}{\true})$ therefore $\conf{P}{\true} \nonewirrbis \conf{Q}{\true}$.
Using the same reasoning we can also show that $\wsatbis \neq \newsymbis$.
\end{example}

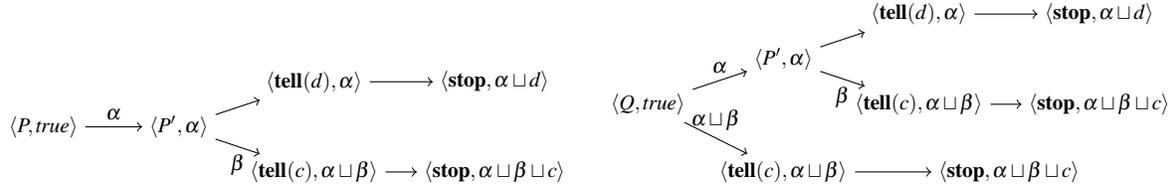
\begin{figure}[tb]
\begin{center}
\begin{tabular}{ll}
\begin{tikzpicture}[scale=0.6]
\scriptsize
\node (A) at (-3, 0) {$\pairccp{P}{\true}$};
\node (B) at (0, 0) {$\pairccp{P'}{\alpha}$};
\node (C) at (3,1) {$\pairccp{\tellp{d}}{\alpha}$};
\node (D) at (3,-1) {$\pairccp{\tellp{c}}{\alpha \sqcup \beta}$};
\node (E) at (7, -1) {$\pairccp{\stopp}{\alpha \sqcup \beta \sqcup  c}$};
\node (F) at (7, 1) {$\pairccp{\stopp}{\alpha \sqcup d}$};
\draw[->] (A) to node[above] {$\alpha$} (B);
\draw[->] (B) to  (C);
\draw[->] (B) to node[below] {$\beta$} (D);
\draw[->] (C) to  (F);
\draw[->] (D) to  (E);
\end{tikzpicture} &
\begin{tikzpicture}[scale=0.6]
\scriptsize
\node (A) at (0, 0) {$\pairccp{Q}{\true}$};
\node (B) at (3, 1) {$\pairccp{P'}{\alpha}$};
\node (C) at (6,2) {$\pairccp{\tellp{d}}{\alpha}$};
\node (D) at (6,-0) {$\pairccp{\tellp{c}}{\alpha \sqcup \beta}$};
\node (E) at (10, 0) {$\pairccp{\stopp}{\alpha \sqcup \beta \sqcup  c}$};
\node (F) at (10, 2) {$\pairccp{\stopp}{\alpha \sqcup d}$};
\node (G) at (3, -1.5) {$\pairccp{\tellp{c}}{\alpha \sqcup \beta}$};
\node (H) at (8, -1.5) {$\pairccp{\stopp}{\alpha \sqcup \beta \sqcup c}$};
\draw[->] (A) to node[above] {$\alpha$} (B);
\draw[->] (B) to  (C);
\draw[->] (B) to node[below] {$\beta$} (D);
\draw[->] (C) to  (F);
\draw[->] (D) to  (E);
\draw[->] (A) to node[above] {$\alpha \sqcup \beta$} (G);
\draw[->] (G) to  (H);
\end{tikzpicture}
\end{tabular}
\caption{Execution of $\conf{P}{\true}$ and $\conf{Q}{\true}$}
\label{fig:milnerProc}
\end{center}
\end{figure}

%

\section{Reducing weak bisimilarity to Strong in CCP} \label{sec:weakCCP}

In this section we shall provide a method for deciding weak bisimilarity in ccp.
As shown in Sec. \ref{ssec:CExMilner}, the usual method for deciding weak bisimilarity
(introduced in Sec. \ref{sssec:weakToStrong}) does not work for ccp.
We shall proceed by redefining $\newtrans{}$ in such a way that it is sound and complete for ccp.
Then we prove that, w.r.t. $\newtrans{}$, symbolic and irredundant bisimilarity
coincide with $\wsatbis$, i.e. $\wsatbis = \newsymbis = \newirrbis$.
We therefore conclude that the partition refinement algorithm in \cite{Aristizabal:12:SAC}
can be used to verify $\wsatbis$ w.r.t. $\newtrans{}$.

\subsection{Defining a new saturation method for CCP} \label{ssec:newLTS}

If we analyze the counter-example to completeness (see Fig. \ref{fig:milnerCompleteness}),
one can see that the problem arises because of the nature of
the labels in ccp, namely using this method $\conf{\askp{\A}{(\askp{\B}{\Stop})}}{\true}$
does not have a transition with $\A \lub \B$ to $\conf{\Stop}{\A \lub \B}$, hence
that fact can be exploited to break the relation among the weak equivalences.
Following this reasoning, instead of only forgetting about the silent actions
we also take into account that labels in ccp can be added together. Thus we
have a new rule that creates a new transition for each two consecutive ones,
whose label is the lub of the labels in them. This method
can also be thought as the reflexive and transitive closure of the
labeled transition relation ($\trans{\A}$). This transition relation
turns out to be sound and complete and it can be used to decide $\wsatbis$.


\subsubsection{A new saturation method} \label{sssec:defNewLTS}
Formally, our new transition relation $\newtrans{}$ is defined by the rules in Tab. \ref{tab:newlabsem}.
For simplicity, we are using the same arrow $\newtrans{}$ to denote this transition relation.
Consequently the definitions of weak barbs, symbolic and irredundant bisimilarity
are now interpreted w.r.t. $\newtrans{}$ ($\genbarb{}, \newsymbis$ and $\newirrbis$ respectively).
\begin{table}[tb]
\small
\[\begin{array}{|c|}
\hline
\makebox{\rTau} \quad \bigfrac{}{\G \newtrans{} \G} \qquad
\makebox{\rLabel} \quad \bigfrac{\G \trans{\A} \G'}{\G \newtrans{\A} \G'} \qquad
\makebox{\rAdd} \quad \bigfrac{\G \newtrans{\A} \G' \newtrans{\B} \G''}
{\G \newtrans{\A \lub \B} \G''} \\
\hline
\end{array}\]
\caption{New Labelled Transition System.}\label{tab:newlabsem}
\end{table}

First, $\genbarb{}$ coincides with $\wbarb{}$, since
a transition in $\newtrans{}$ corresponds to a sequence of reductions.
\begin{lemma}
\label{lem:redsToNewTrans}
$\G \reds \G'$ iff $\G \newtrans{} \G'$.
\end{lemma}

Using this lemma, it is straightforward to see that the notions of weak barbs coincide.
\begin{proposition}
\label{prop:weakBarbs}
$\G \wbarb{e}$ iff $\G \genbarb{e}$.
\end{proposition}

An important property is that the new labeled transition system ($\newtrans{}$) is finitely branching.
Under the assumption that the transition relation $\trans{}$ is finitely branching and that the amount of states
in the transition system is finite,
this way, we can use the fact that labels in ccp are idempotent to prove that $\newtrans{}$ is finitely branching.
Formally:

\begin{proposition}
\label{prop:finBranch}
If for any $\G$ we have $|\{ (\G', \A) | \exists \A. \G \trans{\A} \G'\}| < \infty$ and
$|\{ \G' | \exists \A_1, \dots, \A_n. \G \trans{\A_1} \dots \trans{\A_n} \G'\}| < \infty$,
then $|\{ (\G', \A) | \exists \A. \G \newtrans{\A} \G'\}| < \infty$.

\end{proposition}

\subsubsection{Soundness and Completeness} \label{sssec:soundCompleteNewLTS}
As mentioned before, soundness and completeness of the relation
are the core properties when proving $\satstbisim = \symbis = \irrbis$.
We now proceed to show that our method enjoys of these properties and they
will allow us to prove the correspondence among the equivalences for the weak case.

\begin{lemma}[Soundness of $\newtrans{}$]
\label{lem:soundnessNew}
If $\newtransition{P}{c}{\A}{P'}{c'}$ then $\newtransition{P}{c \lub \A}{}{P'}{c'}$.
\begin{proof}
We proceed by induction on the depth of the inference of $\newtransition{P}{c}{\A}{P'}{c'}$.
\begin{itemize}
 \item Using \rTau we have $\newtransition{P}{c}{}{P}{c}$ and the result follows directly given that $\A = \true$.
 \item Using \rLabel we have $\newtransition{P}{c}{\A}{P'}{c'}$ then $\transition{P}{c}{\A}{P'}{c'}$. By Lemma \ref{lem:correctness} (soundness of $\trans{}$) we get $\transition{P}{c \lub \A}{}{P'}{c'}$ and finally by rule \rLabel $\newtransition{P}{c \lub \A}{}{P'}{c'}$.
 \item Using \rAdd then we have $\newtransition{P}{c}{\B \lub \C}{P'}{c'}$ then $\newtransition{P}{c}{\B}{P''}{c''} \newtrans{\C} \conf{P'}{c'}$ where $\B \lub \C = \A$. By induction hypothesis, $\newtransition{P}{c \lub \B}{}{P''}{c''}$ (1) and $\newtransition{P''}{c'' \lub \C}{}{P'}{c'}$ (2). By monotonicity on (1), $\newtransition{P}{c \lub \B \lub \C}{}{P''}{c'' \lub \C}$ and by rule \rAdd on this transition and (2) then, given that $\B \lub \C = \A$, we obtain $\newtransition{P}{c \lub \A}{}{P'}{c'}$.
\end{itemize}
\end{proof}
\end{lemma}

\begin{lemma}[Completeness of $\newtrans{}$]
\label{lem:completenessNew}
If $\newtransition{P}{c \lub a}{}{P'}{c'}$ then there exist $\A$ and $b$ s.t. $\newtransition{P}{c}{\A}{P'}{c''}$ where $\A \lub b = a$ and $c'' \lub b = c'$.
\begin{proof}
Assuming that $\newtransition{P}{c \lub a}{}{P'}{c'}$ then,
from Lemma \ref{lem:redsToNewTrans},
we can say that $\conf{P}{c \lub a} \reds
\conf{P'}{c'}$ which can be written as $\conf{P}{c \lub a} \trans{} 
\dots \trans{} \conf{P_i}{c_i} \trans{} \conf{P'}{c'}$, we will proceed by
induction on $i$.
\begin{itemize}
 \item [(Base Case)] Assuming $i=0$ then $\transition{P}{c \lub a}{}{P'}{c'}$ and the
result follows directly from Lemma \ref{lem:correctness} (Completeness of $\trans{}$) and
\rLabel.
 \item [(Induction)] Let us assume that $\conf{P}{c \lub a} \trans{}^i \conf{P_i}{c_i}
\trans{} \conf{P'}{c'}$ then by induction hypothesis there exist $\B$ and $b'$ s.t.
$\newtransition{P}{c}{\B}{P_i}{c'_i}$ (1) where $\B \lub b' = a$ and $c'_i \lub b' =
c_i$. Now by completeness on the last transition $\transition{P_i}
{\overbrace{c_i}^{c'_i \lub b'}}{}{P'}{c'}$, there exists $\C$ and $b''$ s.t.
$\transition{P_i}{c'_i}{\C}{P'}{c''}$ where $\C \lub b'' = b'$ and $c'' \lub b'' =
c'$, thus by rule \rLabel we have $\newtransition{P_i}{c'_i}{\C}{P'}{c''}$ (2). We can
now proceed to apply rule \rAdd on (1) and (2) to obtain the transition
$\newtransition{P}{c}{\A}{P'}{c''}$ where $\A = \B \lub \C$ and finally take $b =
b''$, therefore the conditions hold $\A \lub b = \B \lub \C \lub b''= a$ and $c'' \lub
b = c'' \lub b'' = c'$.
\end{itemize}
\end{proof}
\end{lemma}

\subsection{Weak saturated bisimilarity coincides with the strong symbolic and irredundant bisimilarity} \label{ssec:weakToStrongCCP}
We show our main result, a method for deciding $\wsatbis$.
Recall that $\wsatbis$ is the standard weak bisimilarity for ccp \cite{Aristizabal:11:FOSSACS},
and it is defined in terms of $\trans{}$, therefore it does not depend on $\newtrans{}$.
Roughly, we start from the fact that ccp-PR is able to check whether
two configurations are irredundant bisimilar $\irrbis$. Such configurations evolve
according to a transition relation ($\trans{}$), then we provide a new way for them
to evolve ($\newtrans{}$) and we use the same algorithm to compute now $\newirrbis$.
Here we prove that $\wsatbis = \newsymbis = \newirrbis$ hence we give
a reduction from $\wsatbis$ to $\newirrbis$ which has an effective decision procedure.

Given that the transition relation $\trans{}$ (see Lemma \ref{lem:correctness}) is sound and complete,
the correspondence between the symbolic and irredundant bisimilarity follows from \cite{Aristizabal:12:SAC}.
\begin{corollary}
\label{cor:newSymEqNewIrr}
$\G \; \newsymbis \; \G'$ iff $\G \; \newirrbis \; \G'$
\end{corollary}

Finally, in the next two lemmata, we prove that $\wsatbis = \newsymbis$.
\begin{lemma}
\label{lem:weakToNewSym}
If $\G \; \wsatbis \; \G'$ then $\G \; \newsymbis \; \G'$
\begin{proof}
We need to prove that $\R = \{ (\conf{P}{c}, \conf{Q}{d}) \ |\ \conf{P}{c} \wsatbis
\conf{Q}{d}\}$ is a symbolic bisimulation over $\newtrans{}$. The first condition (i)
of the bisimulation follows directly from Proposition \ref{prop:weakBarbs}. As for
(ii), let us assume that $\newtransition{P}{c}{\A}{P'}{c'}$ then by soundness of
$\newtrans{}$ we have $\newtransition{P}{c \lub \A}{}{P'}{c'}$, now by Lemma
\ref{lem:redsToNewTrans} we obtain $\conf{P}{c \lub \A} \reds \conf{P'}{c'}$. Given
that $\conf{P}{c} \wsatbis \conf{Q}{d}$ then from the latter transition we can
conclude that $\conf{Q}{d \lub \A} \reds \conf{Q'}{d'}$ where $\conf{P'}{c'} \wsatbis
\conf{Q'}{d'}$, hence we can use Lemma \ref{lem:redsToNewTrans} again to deduce that
$\newtransition{Q}{d \lub \A}{}{Q'}{d'}$. Finally, by completeness of $\newtrans{}$,
there exist $\B$ and $b$ s.t. $t = \newtransition{Q}{d}{\B}{Q'}{d''}$ where $\B \lub b
= \A$ and $d'' \lub b = d'$, therefore $t \deriv \newtransition{Q}{d}{\A}{Q'}{d'}$ and
$\rel{P'}{c'}{Q'}{d'}$.
\end{proof}
\end{lemma}

\begin{lemma}
\label{lem:newSymToWeak}
If $\G \; \newsymbis \; \G'$ then $\G \; \wsatbis \; \G'$
\begin{proof}
 We need to prove that $\R = \{ (\conf{P}{c \lub a}, \conf{Q}{d \lub a}) \ |\ \conf{P}
{c} \newsymbis \conf{Q}{d}\}$ is a weak saturated bisimulation. First, condition (i)
follows form Proposition \ref{prop:weakBarbs} and (iii) by definition of $\R$. Let us
prove condition (ii), assume $\conf{P}{c \lub a} \reds \conf{P'}{c'}$ then by Lemma \ref{lem:redsToNewTrans}
$\newtransition{P}{c \lub a}{}{P'}{c'}$.
Now by completeness of $\newtrans{}$ there exist $\A$ and $b$ s.t.
$\newtransition{P}{c}{\A}{P'}{c''}$ where $\A \lub b = a$ and $c'' \lub b = c'$.
Since $\conf{P}{c} \newsymbis \conf{Q}{d}$ then we know there exists a
transition $t = \newtransition{Q}{d}{\B}{Q'}{d'}$ s.t. $t \deriv \newtransition{Q}{d}
{\A}{Q'}{d''}$ and $\mathbf{(a)} \conf{P'}{c''} \newsymbis \conf{Q'}{d''}$, by definition of
$\deriv$ there exists $b'$ s.t. $\B \lub b' = \A$ and $d' \lub b' = d''$.
Using soundness of $\newtrans{}$ on $t$ we get $\newtransition{Q}{d
\lub \B}{}{Q'}{d'}$, thus by Lemma \ref{lem:redsToNewTrans} $\conf{Q}{d \lub \B} \reds
\conf{Q'}{d'}$ and finally by monotonicity
\begin{equation}
\label{eq:newSymToWeak:1}
\conf{Q}{d \lub \overbrace{\underbrace{\B \lub b'}_\A \lub b}^a} \reds \conf{Q'}{\overbrace{d' \lub b'}^{d''} \lub b}
\end{equation}
Then, the transition $\redstransition{P}{c \lub a}{P'}{c'}$ can be
rewritten as $\redstransition{P}{c \lub a}{P'}{c'' \lub b}$, and using
\eqref{eq:newSymToWeak:1}, $\redstransition{Q}{d \lub a}{Q'}{d'' \lub b}$. It is left to prove
that $\rel{P'}{c'' \lub b}{Q'}{d'' \lub b}$ which follows from $\mathbf{(a)}$.
\end{proof}
\end{lemma}

Using Lemma \ref{lem:weakToNewSym} and Lemma \ref{lem:newSymToWeak} we obtain
the following theorem.

\begin{theorem}
\label{th:weakSimEqNewSym}
$\conf{P}{c} \newsymbis \conf{Q}{d}$ iff $\conf{P}{c} \wsatbis \conf{Q}{d}$
\end{theorem}

From the above results, we conclude that $\wsatbis = \newirrbis$.
Therefore, given that using ccp-PR in combination with $\newtrans{}$ (and $\wbarb{}$)
we can decide $\newirrbis$, then we can use the same procedure to check
whether two configurations are in $\wsatbis$.

\section{Concluding Remarks} \label{sec:conclusions}
We showed that the transition relation given by Milner's saturation method is
not complete for ccp (in the sense of Definition \ref{def:completeness}). As
consequence we also showed that weak saturated barbed bisimilarity $\wsatbis$
\cite{Aristizabal:11:FOSSACS}  cannot be computed using the ccp partition
refinement algorithm for (strong) bisimilarity ccp wrt to this transition
relation.
We then presented a new transition relation using another saturation mechanism
and showed that it is complete for ccp. We also showed that
the ccp partition refinement can be used to compute $\wsatbis$ using the new
transition relation.
To the best of our knowledge, this is the first approach to verifying weak
bisimilarity for ccp.  As future work,  we plan to investigate
other calculi where the nature of their transitions systems give rise to similar
situations regarding weak and strong bisimilarity,
in particular timed ccp (tcc) \cite{Saraswat:94:LICS},
non-deterministic timed ccp (ntcc) \cite{Palamidessi:01:CP},
universal temporal ccp (utcc) \cite{Olarte:08:SAC}
and Epistemic ccp (eccp) \cite{eccp-extended-version}.

\bibliographystyle{eptcs}
\bibliography{lab}
\end{document}